\documentstyle[12pt,oldlfont]{article}

\begin{document}

\newcommand{\beq}{\begin{eqnarray}}
\newcommand{\eeq}{\end{eqnarray}}
\newcommand{\nn}{\noindent}
\newcommand{\non}{\nonumber}
\newcommand{\ee}{\mbox{$e^+ e^-$}}
\newcommand{\ra}{\rightarrow}
\newcommand{\s}{\\ \vspace*{-3mm} }

\nn \hspace*{11cm} UdeM-LPN-TH-93-159 \\

\vspace*{0.3cm}

\centerline{\large{\bf SIGNALS OF NEW LEPTONS}}

\vspace*{5mm}

\centerline{\sc G.~Azuelos$^{1,2}$ and A.~Djouadi$^1$}

\vspace*{8mm}

\centerline{$^1$ Laboratoire de Physique Nucl\'eaire, Universit\'e de
Montr\'eal, Canada.}
\centerline{$^2$ TRIUMF, Vancouver, Canada.}

\vspace*{0.1cm}

\begin{small}
\begin{center}
\parbox{14cm}
{\begin{center} ABSTRACT \end{center}

\nn We discuss the production of heavy leptons at a future high--energy
e$^+$e$^-$ linear collider with a center of mass energy of 500 GeV. Signals and
backgrounds are analyzed for their single production in association with
ordinary leptons.}
\end{center}
\end{small}

\vspace*{1cm}

\nn {\bf 1.~Introduction} \\

\vspace*{-2mm}

\nn Many theories beyond the Standard Model (SM) lead to the existence of new
fermions. In most of the cases these new fermions have non--canonical SU(2)$
\times$ U(1) quantum numbers, e.g. the left--handed (LH) components are in
weak isosinglets and/or the right--handed (RH) ones in weak isodoublets.
Examples of such fermions [besides a fourth generation with a heavy neutrino]
are the following [1,2]:  \\
$i)$  In the unifying group SO(10), which is one of the simplest
and most economic extensions of	the SM,	the smallest anomaly--free
fermion	representation has dimension 16.  It contains a	right--handed
neutrino in addition to	the 15 Weyl fermions in	one lepton--quark
generation.  These heavy neutrinos are of the Majorana type. \\
$ii)$ In the	exceptional group E$_6$, which is suggested as
low energy limit of superstring	theories, each generation of fermions
lies in	the 27 dimensional representation.  Thus, in addition to the
usual chiral fields, twelve new	fields are needed to complete the
representation. Among	these, there will be two weak
isodoublets of heavy leptons, a	RH and a LH doublet, and two
isosinglets of neutrinos which can be either of	Dirac or Majorana
type. \\
$iii)$ Mirror fermions, whose chiral properties are opposite to those
of ordinary fermions [i.e. the RH components in isodoublets
and the	LH ones	in isosinglets] appear in many extensions of the
SM. They provide a way to restore left--right
symmetry at the	scale of the electroweak symmetry breaking, and	they
naturally occur	in lattice gauge theories. \s

\nn These new fermions will mix with the ordinary fermions of the SM, and this
mixing will give rise to new currents which determine to a large extent their
decay properties and allow for new production mechanisms [2,3]. In this report,
we briefly discuss the signals and backgrounds for the production of the new
leptons at a high--energy $\ee$ linear collider with a c.m. energy of
500 GeV [4]. A more complete discussion can be found in [3]. \\

\nn {\bf 2.~Production and Decay} \\

\vspace*{-2mm}

\nn If the new particles have non--zero electromagnetic and weak charges, they
can be pair produced if their masses are smaller than the beam energy.
In general the reactions are built--up by a superposition of photon and $Z$
boson exchanges [additional contributions could come from extra
gauge bosons if their masses are not much larger than the c.m.~energy of the
collider]. The cross sections are large and, up to phase space suppression
factors, of the order of the point-like QED cross section for muon
pair production. At a 500 GeV $\ee$ collider this gives cross sections
$\sigma \sim 400$ fb which, with the expected integrated luminosity
of $\sim$ 20 fb$^{-1}$ [4], lead to several thousand events. This large
number of
events allows to probe masses up to the kinematical limit of 250 GeV and, due
to the clean environment of $\ee$ colliders, to investigate in detail the
properties of these fermions. \s

\nn The mixing allows an additional production mechanism for the new
fermions: single production in association with their light partners. In the
case of quarks [and for second and third generation new leptons if
inter--generational mixing is neglected] the production process is mediated by
$s$--channel gauge boson exchange; since the mixing angles are restricted to
values smaller than ${\cal O} (10^{-1})$ by present experimental data [1], the
cross sections are rather small. But in the case of [the first generation]
heavy leptons, additional $t$--channel exchanges, $W$ exchange for neutral
leptons and $Z$ exchange for charged leptons are present, increasing the cross
sections by several orders of magnitude. This results in large production rates
which allow to probe lepton masses close to the total c.m. energy if
the mixing angles are not prohibitively tiny. For instance, for a heavy
neutrino with a LH mixing to the electron, the production cross section
is of ${\cal O}(1~{\rm pb})$ at a 500 GeV collider if the neutrino mass is
around 300 GeV and its mixing angle close to the bound $\theta_{\rm mix} \simeq
0.1$. \s

\nn Present experimental data imply that the masses of the new states are
larger than $M_Z/2$ and for not too small mixing angles, larger than $M_W$; the
mass range up to $m_F \sim M_Z$ can be probed at LEP200. The heavy fermions
decay through mixing into massive gauge bosons plus their ordinary light
partners; for masses larger than $M_W/M_Z$  the vector bosons will be on--shell
and will decay into light quarks and leptons. Because the decay is suppressed
by
small mixing angles, the new fermions have very narrow widths: for $\theta_{\rm
mix} \sim 0.1$ and masses around 100 GeV the partial decay widths are less
than 10 MeV. The charged current decay mode is always dominant and
asymptotically the branching ratios are 1/3 and 2/3 for the neutral and charged
current decays, respectively.  To fully reconstruct the fermions from their
final decay products, one needs the branching ratio of their decay into visible
particles. For the decays of $N,E$ into charged leptons and $W/Z$ bosons which
subsequently decay into two jets, the branching ratios are
approximately 0.43 and 0.23, respectively [in the case of the $Z$ boson, one
can
also include the cleaner $e$ and $\mu$ decays, but the branching ratio is
rather small: $\sim 6\%$ compared to $\sim 70\%$ for hadrons].\s

\nn In the following, we will discuss the signals and the various backgrounds
for the single production of charged and neutral heavy leptons [with a LH
mixing to electrons] at a 500 GeV collider. The analysis will be based on an
integrated luminosity of 50 fb$^{-1}$ which corresponds to $\sim 2$ years
running of a standard \ee\ collider [4]. \\

\nn {\bf 3. Signals and Backgrounds} \\

\vspace*{-2mm}

\nn The processes for the production of the charged and neutral heavy leptons
$E$ and $N$ were simulated by incorporating in the PYTHIA [5] generator the
matrix elements for the three body reactions: $e^+ e^- \ra N \nu_e (E^\pm e^\mp
) \ra \nu_e e^\pm W^\mp (e^\pm e^\mp Z)$ and forcing the gauge bosons to decay
hadronically; full hadronization was allowed to take place. All the resulting
particles were then subjected to detector resolution and acceptance effects.
The parameters for the detector were taken from the ``standard'' set of [6]
but with angular acceptance up to $|\cos\theta| < 0.98$ for the electromagnetic
and hadronic calorimeters as well as for the charged particle tracker. In the
case of the neutral heavy lepton, the missing momentum vector was calculated
and subsequently assumed to be the reconstructed neutrino momentum. The
background processes were simulated using existing parameter options in PYTHIA.
\s

\nn In the case of the charged lepton $E$, the signal consists of an \ee\ pair
and two jets. Other processes likely to produce such a configuration are:
(i) $ e^+ e^- \to e^+ e^- Z$, with the $Z$ decaying hadronically; the cross
section is 3800 fb;
(ii)  $ e^+ e^- \to Z Z$, with a cross section of 615 fb;
(iii) $ e^+ e^- \to t \bar{t} $, followed by  $ t \to b W$ and leading to two
electrons and 2 jets but with missing momentum; and
(iv) $ \gamma \gamma \to e^+ e^- q \bar{q} $  which has a large cross section
but the jets have small invariant masses and the resulting  events have the
primary electrons going mostly along the beam pipe.
In the case of the neutral lepton $N$, the signal consists of an electron, a
pair of jets and missing momentum. The backgrounds that one has to consider
are:
(i) $ \ee\  \to e \nu W$ with a cross section of 5800 fb when the $W$ decays
hadronically;
 (ii) $ \ee\ \to WW$ where one of the $W$'s decays hadronically and the
other to an $e \nu$ pair and the cross section in this case is 1140 fb;
(iii) $ \gamma \gamma \to e (e)  q \bar{q} $ where one of the electrons
escapes observation. \s

\nn The analysis proceeded by selecting among all ``reconstructed'' final state
particles the $e^-, e^+$ with a momentum greater than 30 GeV. The following
cuts were found to suppress considerably the background processes without
affecting significantly the signals. \s

\nn For the charged lepton: (1) One and only one \ee\ pair. (2) $ 85 < M_{jj}
< 105 $ GeV where the lower bound on the reconstructed $Z$ mass was set
intentionally high so as to avoid a possibility of confusing a $Z$ with a $W$.
(3) $|p_{l^+}| > (E_{\rm beam}-M_E^2/4E_{\rm beam})-40$ GeV and $|p_{l^-}| >
\frac{2}{3} M_E -133$ GeV, with $M_E$ the reconstructed mass; these kinematic
constraints ensure energy-momentum conservation, with some tolerance for
detector resolution effects. (4) A cut $|M_{ll} - M_Z| > 12$ GeV which is
effective against the $ZZ$ background. (5) Cuts $\cos\theta_Z > -(M_E+440)/720$
and $\cos\theta_Z < (2100-M_E)/2000$, which are necessary to reduce the
background from $ e^+ e^- \to e^+ e^- Z$. (6) $\cos\theta_{ll} < 0.5$ since in
the signal, the
two leptons are mostly back--to--back. Cuts (3) and (5) apply to $E^-$, similar
ones apply for $E^+$. \s

\nn In the case of neutral leptons, similar cuts where applied: (1) One and
only one $e^-$. (2) $ 70 < M_{jj} < 90 $ GeV for the reconstructed $W$ mass.
(3) $|p_{\nu} | > (E_{\rm beam} -M_N^2/4E_{\rm beam})-40$ and $|p_{l}| >
\frac{2}{3} M_N - 133$. (4) A cut $ M_{ll} > 120$ GeV, effective against the
$WW$ background. (5) $\cos\theta_W < 2.58-M_N/240$ reduces the background from
$ e^+ e^- \to e^- \bar{\nu} W$. (6) $\cos\theta_{l\nu} < 0.5$; here also the
two leptons from the signal are mostly back--to--back. \\

\nn {\bf 4. Results and Discussions} \\

\vspace*{-2mm}

\nn Fig.~1 shows the reconstructed invariant mass histograms for heavy leptons
with masses of 250, 350 and 450 GeV and with mixing angles $\theta_{\rm mix}
=0.05$ in the case of the charged lepton and $\theta_{\rm mix} = 0.025$ for
the heavy neutrino. For these values, one can see that the signal peaks stand
out clearly from the background events, especially for heavy lepton masses not
too close to the total c.m. energy of the collider. For the $\gamma \gamma$ and
$t \bar{t}$ backgrounds, no events survived [in the former case, because it was
not possible to generate a sufficient number of events, only an upper limit of
$< 15$ surviving background events/5 GeV can be given]. The backgrounds from
vector boson pair production have been suppressed to a very low level; the
backgrounds from single $W$ and $Z$ production are relatively higher. \s

\nn For smaller mixing angles, the signal cross sections have to be scaled down
correspondingly. For lepton masses of the order of 450 GeV, only slightly
smaller $\theta_{\rm mix}$ values can be probed, while for masses around 350
GeV one can go down by at least a factor of two. The situation is much more
favorable for heavy neutrinos than for charged leptons, the cross section
being one order of magnitude larger. For instance, assuming a mass of 350 GeV
and requiring that the ratio of the signal events to the square root of the
background events be larger than unity, one can probe mixing angles down to
$\theta_{\rm mix} \sim 0.005$ for neutral leptons and $\theta_{\rm mix} \sim
0.03=2$ for charged leptons. \s

\nn In conclusion. Because the cross sections are large and the environment in
\ee\ colliders is very clean, pair produced new leptons can be unambiguously
detected up to masses close to the beam energy. For new leptons of the first
generation, one can reach masses close to the total center of mass energy of
the machine in the process of single production in association with ordinary
electrons and neutrinos, provided that the mixing angles are not prohibitively
tiny. \\

\nn {\bf Acknowledgment:} One of us (A.D.) would like to thank the organizers
of this workshop for creating a nice and very stimulating atmosphere. \\

\nn {\bf References}. \\

\vspace*{-2mm}

\nn [1] See e.g. the talks given by J. Hewett and E. Nardi, these proceedings.
\s

\nn [2] See A.~Djouadi, D.~Schaile and C. Verzegnassi [conv.] et al., Report
of the Working \\ \hspace*{5mm} Group ``Extended Gauge Models", Proceedings of
the Workshop ``\ee\ Collisions \\ \hspace*{5mm} at 500 GeV: The Physics
Potential", Report DESY 92-123B, P.~Zerwas, ed. \s

\nn [3] For details and for a complete set of references, see A.~Djouadi, Prep.
UdeM--LPN--TH \\ \hspace*{5mm} --93--157, and G.~Azuelos and A. Djouadi, Prep.
UdeM--LPN--TH--93--158. \s

\nn [4] See e.g. the talk given by B. Wiik, these proceedings. \s

\nn [5] T. Sj\"ostrand, PYTHIA 5.6 and JETSET 7.3, Report CERN-TH.6488/92. \s

\nn [6] P. Grosse-Wiesmann, D. Haidt and J. Schreiber, in the same proceedings
as in Ref.~[2].

\end {document}